\documentclass{article}
\usepackage{amssymb}
\usepackage{amsmath}
\usepackage{harvard}

\newtheorem{theorem}{Theorem}
\newtheorem{acknowledgement}[theorem]{Acknowledgement}

\input{tcilatex}

\begin{document}

\title{Einstein-Yang-Mills black hole solution in higher dimensions by the
Wu-Yang Ansatz}
\author{S. Habib Mazharimousavi$^{\ast }$ and M. Halilsoy$^{\dagger }$ \\
%EndAName
Department of Physics, Eastern Mediterranean University,\\
G. Magusa, North Cyprus, Mersin-10, Turkey\\
$^{\ast }$habib.mazhari@emu.edu.tr\\
$^{\dagger }$mustafa.halilsoy@emu.edu.tr}
\maketitle

\begin{abstract}
By employing the higher ($N\geq 5$) dimensional version of the Wu-Yang
Ansatz we obtain black hole solutions in the spherically symmetric
Einstein-Yang-Mills (EYM) theory. Although these solutions were found
recently by other means, our method provides an alternative way in which one
identifies the contribution from the Yang-Mills (YM) charge. Our method has
the advantage to be carried out analytically as well. We discuss some
interesting features of the black hole solutions obtained.
\end{abstract}

\section{\protect\bigskip Introduction\protect\bigskip\ }

Black holes in Einstein and Einstein-Maxwell (EM) theory have a long
history, starting with Schwarzschild as early as 1916 which is a well-known
subject by now. Higher dimensional version of black holes has also attracted
attention during the last three decades [1,2]. Extension of this fascinating
subject to the Einstein-Yang-Mills (EYM) theory is relatively rather
new[3-5]. This originates from the intricate structure of the YM system
compared with the Maxwell's electromagnetism. Finding exact solutions to the
YM system in a flat background is a challenging problem by itself, not to
mention its coupled form with gravity. In spite of all its inherent
complications when specified to spherical symmetry and specific gauge group
such as $SO(N-1)$ analytical solutions can be obtained. From this token
black hole solutions have been obtained newly[4-7].

Our aim in this Letter is to show that black hole solutions can be obtained
by an alternative method, namely by making use of the Wu-Yang Ansatz [8-9]
in higher $(N>4)$ dimensions. Our method has the advantage of introducing
the YM charge ab initio and obtain, after integrating the EYM equations, the
charge dependent term in the metric function. Except for the $N=5$ case, the
EYM equations dictate a Reissner-Nordstrom type black hole solution. The
case $N=5,$ however, leads to a black hole which incorporates a logaritmic
term unprecedented in the $N>5$ cases. For this reason we shall treat the
case $N=5$ in some detail while $N>5$ cases will be considered separately.

\section{Field equations in Einstein-Yang-Mills theory}

The action which describes Einstein-Yang-Mills gravity without a
cosmological constant in $N$ dimensions reads%
\begin{equation}
I_{G}=\frac{1}{2}\int_{\mathcal{M}}dx^{N}\sqrt{-g}\left[ R-\overset{\left(
N-1\right) (N-2)/2}{\underset{a=1}{\tsum }}F_{\mu \nu }^{(a)}F^{(a)\mu \nu }%
\right]
\end{equation}%
where $R$ is the Ricci Scalar and the YM fields $F_{\mu \nu }^{\left(
a\right) }$ are 
\begin{equation}
F_{\mu \nu }^{\left( a\right) }=\partial _{\mu }A_{\nu }^{\left( a\right)
}-\partial _{\nu }A_{\mu }^{\left( a\right) }+\frac{1}{2\sigma }C_{\left(
b\right) \left( c\right) }^{\left( a\right) }A_{\mu }^{\left( b\right)
}A_{\nu }^{\left( c\right) }
\end{equation}%
where $C_{\left( b\right) \left( c\right) }^{\left( a\right) }$ are the
structure constants of $\frac{\left( N-1\right) (N-2)}{2}$-parameter Lie
group $G,$ $\sigma $ is a coupling constant, and $A_{\mu }^{\left( a\right)
} $ are the gauge potentials. We note that the internal indices $%
\{a,b,c,...\}$ do not differ whether in covariant or contravariant form.
Variation of the action with respect to the space-time metric $g_{\mu \nu }$
yields the EYM equations%
\begin{equation}
G_{\mu \nu }=T_{\mu \nu },
\end{equation}%
where the stress-energy tensor is 
\begin{equation}
T_{\mu \nu }=\overset{\left( N-1\right) (N-2)/2}{\underset{a=1}{\tsum }}%
\left[ 2F_{\mu }^{\left( a\right) \lambda }F_{\nu \lambda }^{\left( a\right)
}-\frac{1}{2}F_{\lambda \sigma }^{\left( a\right) }F^{\left( a\right)
\lambda \sigma }g_{\mu \nu }\right] .
\end{equation}

Variation of the action with respect to the gauge potentials $A_{\mu
}^{\left( a\right) }$ yields the Yang-Mills equations 
\begin{equation}
F_{;\mu }^{\left( a\right) \mu \nu }+\frac{1}{\sigma }C_{\left( b\right)
\left( c\right) }^{\left( a\right) }A_{\mu }^{\left( b\right) }F^{\left(
c\right) \mu \nu }=0
\end{equation}%
while the integrability conditions are 
\begin{equation}
\ast F_{;\mu }^{\left( a\right) \mu \nu }+\frac{1}{\sigma }C_{\left(
b\right) \left( c\right) }^{\left( a\right) }A_{\mu }^{\left( b\right) }\ast
F^{\left( c\right) \mu \nu }=0
\end{equation}%
in which * means duality [10].

\section{5D EYM black holes}

Unlike the more general form of the 5-dimensional spherically symmetric line
element, given by Brihaye et al [4], we restrict ourself to work with a
symmetric version of it [6], to gain the benefits of the Wu-Yang Ansatz in
Yang-Mills equations (i.e., this symmetric Ansatz provides us to cancel the
role of gravity in YM equations, which means that, the metric function $f(r)$
will not appear in the YM equations) as%
\begin{equation}
ds^{2}=-f(r)\;dt^{2}+\frac{dr^{2}}{f(r)}+r^{2}d\;\Omega _{3}^{2}
\end{equation}%
where 
\begin{equation}
d\Omega _{3}^{2}=d\theta ^{2}+\sin ^{2}\theta d\phi ^{2}+\sin ^{2}\theta
\sin ^{2}\phi \ d\psi ^{2}
\end{equation}%
in which 
\begin{equation*}
0\leq \theta \leq \pi ,0\leq \phi ,\psi \leq 2\pi .
\end{equation*}

We introduce the new 5-dimentional Wu-Yang Ansatz [11] as follows%
\begin{eqnarray}
A^{(a)} &=&\frac{Q}{r^{2}}\left( x_{i}dx_{j}-x_{j}dx_{i}\right) \\
2 &\leq &i\leq 4,  \notag \\
1 &\leq &j\leq i-1  \notag \\
1 &\leq &\left( a\right) \leq 6  \notag
\end{eqnarray}%
or explicitly%
\begin{equation}
\begin{array}{c}
A^{\left( 1\right) }=\frac{Q}{r^{2}}\left( x_{2}dx_{1}-x_{1}dx_{2}\right) \\ 
A^{\left( 2\right) }=\frac{Q}{r^{2}}\left( x_{3}dx_{1}-x_{1}dx_{3}\right) \\ 
A^{\left( 3\right) }=\frac{Q}{r^{2}}\left( x_{3}dx_{2}-x_{2}dx_{3}\right) \\ 
A^{\left( 4\right) }=\frac{Q}{r^{2}}\left( x_{4}dx_{1}-x_{1}dx_{4}\right) \\ 
A^{\left( 5\right) }=\frac{Q}{r^{2}}\left( x_{4}dx_{2}-x_{2}dx_{4}\right) \\ 
A^{\left( 6\right) }=\frac{Q}{r^{2}}\left( x_{4}dx_{3}-x_{3}dx_{4}\right)%
\end{array}%
\end{equation}%
where $Q$ is the only non-zero gauge charge and%
\begin{eqnarray}
x_{1} &=&r\cos \psi \sin \phi \sin \theta \\
x_{2} &=&r\sin \psi \sin \phi \sin \theta  \notag \\
x_{3} &=&r\cos \phi \sin \theta  \notag \\
x_{4} &=&r\cos \theta .  \notag
\end{eqnarray}

By setting $\sigma =Q$ in Eq. (2), the YM field 2-forms are given by%
\begin{equation}
F^{\left( a\right) }=dA^{\left( a\right) }+\frac{1}{2Q}C_{\left( b\right)
\left( c\right) }^{\left( a\right) }A^{\left( b\right) }\wedge A^{\left(
c\right) }.
\end{equation}

We note that our notation follows the standard exterior differential forms,
namely $d$ stands for the exterior derivative, while $\wedge $ stands for
the wedge product. The hodge star * in the sequel will be used to represent
duality [10].

The YM integrability conditions

\begin{equation}
dF^{\left( a\right) }+\frac{1}{Q}C_{\left( b\right) \left( c\right)
}^{\left( a\right) }A^{\left( b\right) }\wedge F^{\left( c\right) }=0
\end{equation}%
are easily satisfied by using $\left( 10\right) .$ The YM equations 
\begin{equation}
d\ast F^{\left( a\right) }+\frac{1}{Q}C_{\left( b\right) \left( c\right)
}^{\left( a\right) }A^{\left( b\right) }\wedge \ast F^{\left( c\right) }=0
\end{equation}%
are all satisfied as well.

The energy-momentum tensor 
\begin{equation}
T_{\mu \nu }=\overset{6}{\underset{a=1}{\tsum }}\left[ 2F_{\mu }^{\left(
a\right) \lambda }F_{\nu \lambda }^{\left( a\right) }-\frac{1}{2}F_{\lambda
\sigma }^{\left( a\right) }F^{\left( a\right) \lambda \sigma }g_{\mu \nu }%
\right]
\end{equation}%
where $\overset{6}{\underset{a=1}{\tsum }}\left[ F_{\lambda \sigma }^{\left(
a\right) }F^{\left( a\right) \lambda \sigma }\right] =6Q^{2}/r^{4}$, has the
non-zero components%
\begin{eqnarray}
T_{tt} &=&\frac{3Q^{2}f(r)}{r^{4}} \\
T_{rr} &=&-\frac{3Q^{2}}{r^{4}f(r)}  \notag \\
T_{\theta \theta } &=&\frac{Q^{2}}{r^{2}}  \notag \\
T_{\phi \phi } &=&\frac{Q^{2}\sin ^{2}\theta }{r^{2}}  \notag \\
T_{\psi \psi } &=&\frac{Q^{2}\sin ^{2}\theta \sin ^{2}\phi }{r^{2}}  \notag
\end{eqnarray}

The EYM equations $G_{\mu \nu }=T_{\mu \nu },$ reduce to the simple set of
equations%
\begin{eqnarray}
rf^{\prime }\left( r\right) +2\left( f\left( r\right) -1\right) &=&\frac{%
-2Q^{2}}{r^{2}} \\
r^{2}f^{\prime \prime }\left( r\right) +4rf^{\prime }\left( r\right)
+2\left( f\left( r\right) -1\right) &=&\frac{2Q^{2}}{r^{2}}  \notag
\end{eqnarray}%
in which a prime denotes derivative with respect to $r$.

This set admits the solution 
\begin{equation}
f\left( r\right) =1-\frac{m}{r^{2}}-\frac{2Q^{2}}{r^{2}}\ln \left( r\right)
\end{equation}%
in which $m$ is the usual integration constant to be identified as mass
(this solution was obtained by Brihaye et al in different manner in
reference [4]) . The radii of Cauchy horizon $r_{-}$ and the event horizon $%
r_{+}$ are determined from the roots of $f\left( r\right) =0.$ For $m$ and $%
Q\ $chosen outside of the unmarked region $\mathcal{R}$ in the Fig. (1), one
may get%
\begin{eqnarray}
r_{-} &=&\exp \left[ -\frac{1}{2Q^{2}}\left( m+Q^{2}LambertW\left( 0,-\frac{%
e^{-\frac{m}{Q^{2}}}}{Q^{2}}\right) \right) \right] \\
r_{+} &=&\exp \left[ -\frac{1}{2Q^{2}}\left( m+Q^{2}LambertW\left( -1,-\frac{%
e^{-\frac{m}{Q^{2}}}}{Q^{2}}\right) \right) \right]
\end{eqnarray}%
in which $LambertW(k,x)$ is the Lambert function[12]. The asymptotic
behavior of $f\left( r\right) ,$ which admits 
\begin{eqnarray}
\underset{r\rightarrow 0^{+}}{\lim }f\left( r\right) &=&+\infty \\
\underset{r\rightarrow \infty }{\lim }f\left( r\right) &=&1  \notag
\end{eqnarray}%
also and having roots in between reconfirms the existence of $r_{+}$ and $%
r_{-}$ simultaneously. Fig. (1) clearly shows the possible black-hole region
in $m-Q$ plane. In the Fig. (2) we give the radii of the horizons in term of 
$m$ while $Q$ is set to be fixed, and in the Fig. (3) same thing is given
but in term of $Q$, while $m$ is fixed. These two figures clearly show that,
with a large value of charge, the radii of the horizons are m-independent
which is a consequence of logaritmic term in the metric function.

The surface gravity, $\kappa $ defined by [13]%
\begin{equation}
\kappa ^{2}=-\frac{1}{4}g^{tt}g^{ij}g_{tt,i}\;g_{tt,j}
\end{equation}%
has the value%
\begin{equation}
\kappa =\left\vert \frac{1}{2}f^{\prime }\left( r_{+}\right) \right\vert
=\left\vert \frac{1}{r_{+}^{3}}\left( m-Q^{2}+2Q^{2}\ln r_{+}\right)
\right\vert .
\end{equation}

The associated Hawking temperature is given by 
\begin{equation}
T_{H}=\frac{\kappa }{2\pi }=\left\vert \frac{1}{2\pi r_{+}^{3}}\left(
m-Q^{2}+2Q^{2}\ln r_{+}\right) \right\vert
\end{equation}%
in the choice of units $c=G=\hbar =k=1.$

\bigskip

\section{The Wu-Yang Ansatz in $N>5$ dimensions\protect\bigskip \protect%
\bigskip \protect\bigskip}

The N-dimensional line element is chosen as%
\begin{equation}
ds^{2}=-f(r)\;dt^{2}+\frac{dr^{2}}{f(r)}+r^{2}d\;\Omega _{N-2}^{2}
\end{equation}%
in which the $S^{N-2}$ line element will be expressed in the standard
spherical form%
\begin{equation}
d\Omega _{N-2}^{2}=d\theta _{1}^{2}+\underset{i=2}{\overset{N-3}{\tsum }}%
\underset{j=1}{\overset{i-1}{\tprod }}\sin ^{2}\theta _{j}\;d\theta _{i}^{2}
\end{equation}%
where 
\begin{equation*}
0\leq \theta _{1}\leq \pi ,0\leq \theta _{i}\leq 2\pi .
\end{equation*}

We introduce the Wu-Yang Ansatz in N-D as%
\begin{eqnarray}
A^{(a)} &=&\frac{Q}{r^{2}}\left( x_{i}dx_{j}-x_{j}dx_{i}\right) \\
2 &\leq &i\leq N-1,  \notag \\
1 &\leq &j\leq i-1  \notag \\
1 &\leq &\left( a\right) \leq \left( N-1\right) (N-2)/2  \notag
\end{eqnarray}%
where we imply (to have a systematic process) that the super indices $a$ is
chosen according to the values of $i$ and $j$ in order. For instance, we
present some of them 
\begin{equation}
\begin{array}{c}
A^{\left( 1\right) }=\frac{Q}{r^{2}}\left( x_{2}dx_{1}-x_{1}dx_{2}\right) \\ 
A^{\left( 2\right) }=\frac{Q}{r^{2}}\left( x_{3}dx_{1}-x_{1}dx_{3}\right) \\ 
A^{\left( 3\right) }=\frac{Q}{r^{2}}\left( x_{3}dx_{2}-x_{2}dx_{3}\right) \\ 
A^{\left( 4\right) }=\frac{Q}{r^{2}}\left( x_{4}dx_{1}-x_{1}dx_{4}\right) \\ 
A^{\left( 5\right) }=\frac{Q}{r^{2}}\left( x_{4}dx_{2}-x_{2}dx_{4}\right) \\ 
A^{\left( 6\right) }=\frac{Q}{r^{2}}\left( x_{4}dx_{3}-x_{3}dx_{4}\right) \\ 
A^{\left( 7\right) }=\frac{Q}{r^{2}}\left( x_{5}dx_{1}-x_{1}dx_{5}\right) \\ 
A^{\left( 8\right) }=\frac{Q}{r^{2}}\left( x_{5}dx_{2}-x_{2}dx_{5}\right) \\ 
A^{\left( 9\right) }=\frac{Q}{r^{2}}\left( x_{5}dx_{3}-x_{3}dx_{5}\right) \\ 
A^{\left( 10\right) }=\frac{Q}{r^{2}}\left( x_{5}dx_{4}-x_{4}dx_{5}\right)
\\ 
...%
\end{array}%
\end{equation}%
in which $r^{2}=\overset{N-1}{\underset{i=1}{\sum }}x_{i}^{2}.$

The YM field 2-forms are defined by the expression%
\begin{equation}
F^{\left( a\right) }=dA^{\left( a\right) }+\frac{1}{2Q}C_{\left( b\right)
\left( c\right) }^{\left( a\right) }A^{\left( b\right) }\wedge A^{\left(
c\right) }.
\end{equation}

In Sec.3 $($ i.e.$N=5),$ we had $a=1...6$. For $N=6$ we have $a=1...10$, and
in general for $N$ we must have $\left( N-1\right) \left( N-2\right) /2$
gauge potentials. The integrability conditions 
\begin{equation}
dF^{\left( a\right) }+\frac{1}{Q}C_{\left( b\right) \left( c\right)
}^{\left( a\right) }A^{\left( b\right) }\wedge F^{\left( c\right) }=0
\end{equation}%
are easily satisfied by using (28). The YM equations 
\begin{equation}
d\ast F^{\left( a\right) }+\frac{1}{Q}C_{\left( b\right) \left( c\right)
}^{\left( a\right) }A^{\left( b\right) }\wedge \ast F^{\left( c\right) }=0
\end{equation}%
also are all satisfied. The energy-momentum tensor (4), becomes after%
\begin{equation}
\overset{\left( N-1\right) \left( N-2\right) /2}{\underset{a=1}{\tsum }}%
\left[ F_{\lambda \sigma }^{\left( a\right) }F^{\left( a\right) \lambda
\sigma }\right] =\frac{\left( N-3\right) \left( N-2\right) Q^{2}}{r^{4}}
\end{equation}%
with the non-zero components%
\begin{eqnarray}
T_{00} &=&\frac{\left( N-3\right) \left( N-2\right) Q^{2}f\left( r\right) }{%
2r^{4}} \\
T_{11} &=&-\frac{\left( N-3\right) \left( N-2\right) Q^{2}}{2r^{4}f\left(
r\right) }  \notag \\
T_{22} &=&-\frac{\left( N-3\right) \left( N-6\right) Q^{2}}{2r^{2}}  \notag
\\
T_{AA} &=&-\frac{\left( N-3\right) \left( N-6\right) Q^{2}}{2r^{2}}\overset{%
A-2}{\underset{i=1}{\tprod }}\sin ^{2}\theta _{i}  \notag \\
2 &<&A\leq N-1.  \notag
\end{eqnarray}

The EYM equations 
\begin{equation}
G_{\mu \nu }=T_{\mu \nu },
\end{equation}%
reduce to the set of general equations 
\begin{gather}
r^{3}f^{\prime }\left( r\right) +\left( N-3\right) r^{2}\left( f\left(
r\right) -1\right) +\left( N-3\right) Q^{2}=0 \\
r^{4}f^{\prime \prime }\left( r\right) +2\left( N-3\right) r^{3}f^{\prime
}\left( r\right) +\left( N-3\right) \left( N-4\right) r^{2}\left( f\left(
r\right) -1\right) +  \notag \\
\left( N-3\right) \left( N-6\right) Q^{2}=0  \notag \\
N\geq 5  \notag
\end{gather}%
in which a prime denotes derivative with respect to $r$.

These equations admit the solution 
\begin{eqnarray}
f\left( r\right) &=&1-\frac{m}{r^{N-3}}-\frac{\left( N-3\right) Q^{2}}{%
\left( N-5\right) r^{2}} \\
N &>&5  \notag
\end{eqnarray}%
in which $m$ is the usual integration constant to be identified as mass. In
the sequel as particular examples we consider the $N=6$ and $N>5$ cases in
general.

\subsection{$N=6$ case}

Equation (36) for $N=6$ implies%
\begin{equation}
f\left( r\right) =1-\frac{m}{r^{3}}-\frac{3Q^{2}}{r^{2}}
\end{equation}%
from which, since we consider $m$ to be positive, one can only find the
radius of the event horizon (i.e., $r_{+}$) by solving the following
depressed cubic equation 
\begin{equation}
r^{3}-3Q^{2}r-m=0
\end{equation}%
in which the solution admits%
\begin{equation}
r_{+}=\frac{1}{2}\sqrt[3]{\Delta }+\frac{2Q^{2}}{\sqrt[3]{\Delta }}
\end{equation}%
where 
\begin{equation}
\Delta =4m+4\sqrt{m^{2}-4Q^{6}}.
\end{equation}

Following these results, we plot the Fig.s (4) and (5) to show how the
radius of event horizon behaves in terms of the mass and the gauge charge.

Nevertheless the asymptotic behaviors of $f(r)$ is given by 
\begin{equation}
\underset{r\rightarrow 0^{+}}{\lim }f(r)=\underset{r\rightarrow 0^{+}}{\lim }%
\left( 1-\frac{m}{r^{3}}\right) =-\infty
\end{equation}%
and%
\begin{equation}
\underset{r\rightarrow \infty }{\lim }f(r)=\underset{r\rightarrow +\infty }{%
\lim }\left( 1-\frac{3Q^{2}}{r^{2}}\right) =1.
\end{equation}

These imply that in $6$-dimensional EYM-black hole, the Cauchy horizon $%
r_{-} $ is not defined. We comment also that if one sets $m=0$ and $Q\neq 0,$
similar to the 5-dimensional Schwarzschild black hole the metric function
takes the form 
\begin{equation}
f\left( r\right) =1-\frac{3Q^{2}}{r^{2}}
\end{equation}%
whose radius of the event horizon is given by%
\begin{equation}
r_{+}=\sqrt{3}Q.
\end{equation}

\subsection{Arbitrary $N>5$ case}

The metric function $f\left( r\right) $ in any arbitrary dimensions N given
by (36), has the following limits: 
\begin{eqnarray}
\underset{r\rightarrow \infty }{\lim }f(r) &=&1 \\
\underset{r\rightarrow 0^{+}}{\lim }f(r) &=&-\infty .  \notag
\end{eqnarray}

The radius of the event horizon may be determined by finding the root(s) of
the metric function $f\left( r\right) .$ It is not difficult to show that,
with positive mass and N\TEXTsymbol{>}5, $f\left( r\right) $ has only one
positive real root. This manifests\ itself as an essential difference
between $N=5$ and $N>5$ dimensions in which, the black hole solution in five
dimensions always has the Cauchy and event horizon, while for N\TEXTsymbol{>}%
5, the black hole solutions have only the event horizon.

In N-dimensional-EYM-black hole, when the mass of the black hole is zero
(i.e. $m=0$) we have a gauge charged black hole whose radius of event
horizon is given by 
\begin{equation}
r_{+}=Q\sqrt{\frac{N-3}{N-5}}
\end{equation}%
and it is a kind of 5-dimensional Schwarzschild black hole with a metric
function as follows%
\begin{equation}
f\left( r\right) =1-\frac{\left( N-3\right) Q^{2}}{\left( N-5\right) r^{2}}.
\end{equation}

Unlike the case of $N=5$ the EYM black hole for $N>5$ is reminiscent of the
Reissner-Nordstrom black hole in which the mass and charge terms are
naturally separated. However there is a striking difference between EYM
black hole in higher dimensions and the Reissner-Nordstrom, namely the gauge
charge term in $f\left( r\right) $ has the fixed power of $r$ given by $%
1/r^{2}$. The surface gravity has the value 
\begin{eqnarray}
\kappa &=&\left\vert \frac{1}{2}f^{\prime }\left( r_{+}\right) \right\vert
=\left\vert \frac{1}{2}\frac{\left( N-3\right) m}{r_{+}^{N-2}}+\frac{\left(
N-3\right) Q^{2}}{\left( N-5\right) r_{+}^{3}}\right\vert \\
N &>&5  \notag
\end{eqnarray}%
where $r_{+}$ is the radius of the event horizon. The contribution by the
parameters $m,n,r_{+}$\ and $Q\ $\ to the Hawking temperature (i.e. $T_{H}=%
\frac{\kappa }{2\pi }$, with all physical constants $c,$ $\hslash ,$ $G,$ \
and $k$ set to unity) does not require any comment.

\section{Conclusion}

Our method employs the Wu-Yang Ansatz in higher dimensions which can be
handled analytically and where the mass and gauge charge terms are taken
separately. Unlike our approach, in the Ref.s [4-6], a position dependent
mass density $m\left( r\right) $ is assumed which is determined upon
imposition of the EYM equations yielding the mass and constant charge terms.
The EYM black hole has the marked distinction from the Reissner-Nordstrom
black hole in higher dimensions as far as the $r$ dependence of the charge
term is concerned. Another significant difference of N=5 and N\TEXTsymbol{>}%
5 black holes is that in the former case due to the logaritmic term we have
both $r_{+}$ and $r_{-}$ (for possible black holes), whereas for N%
\TEXTsymbol{>}5 we have only $r_{+}.$ The effect of the mass and charge on
black hole formation, the existence of event (Cauchy for N=5) horizons are
plotted in Fig.s [2-5]. We note that the Wu-Yang Ansatz has been used also
in finding black hole solutions in the EYM-Gauss-Bonnet theory [11].

\begin{acknowledgement}
We would like to thank the anonymous referee for the valuable comments and
suggestions.
\end{acknowledgement}

\bigskip 

\bigskip 

\bigskip 

\textbf{Figure captions:}

Figure (1): Eq.s (19)-(20) (in the text) imply that for meaningful $r_{\pm },
$ $m$ and $Q$ must belong to a particular set.

Figure (2): $r_{\pm }$ versus $Q$ plot for different masses according to
Eq.s (19)-(20). For large $Q$ values, with fixed $m$, $r_{+}$ and $r_{-}$ go
to particular limits.

Figure (3): The plot of $r_{\pm }$ versus $m$ for different charges
according to Eq.s (19)-(20). For large $Q$, irrespective of $m$, $r_{+}$ and 
$r_{-}$ converge at constant values.

Figure (4): $r_{+}$ versus $Q$ plot for different masses (from Eq. (39)).
For large $Q$, independent of $m$, it converges to a limit.

Figure (5): $r_{+}$ versus $m$ plot for different charges (from Eq. (39)).
For large $Q$, $r_{+}$ becomes independent of $m$.

\end{document}